\newif\ifjors

\ifjors

\documentclass{jors}

\definecolor{mygray}{gray}{0.6}

\else

\documentclass[conference,a4paper]{IEEEtran}

\usepackage{hyperref}

\fi

\usepackage[official]{eurosym}
\usepackage{flushend,amsmath}

\usepackage[utf8]{inputenc}

  \usepackage[pdftex]{graphicx}
  \graphicspath{{graphics/}}
  \DeclareGraphicsExtensions{.pdf,.jpeg,.png}

\usepackage{tikz}

\usepackage[europeanresistors,americaninductors]{circuitikz}
\usepackage{adjustbox}

\usepackage{fixltx2e}

\usepackage{booktabs}
\newcommand{\ra}[1]{\renewcommand{\arraystretch}{#1}}

\usepackage{pifont}

\usepackage{color}

\def\el{${}_{\textrm{el}}$}
\def\th{${}_{\textrm{th}}$}
\newcommand{\ubar}[1]{\text{\b{$#1$}}}

\newcommand*\rot{\rotatebox{90}}
\newcommand*\OK{\ding{51}}

\begin{document}

\ifjors

{\bf Software paper for submission to the Journal of Open Research Software} \\

To complete this template, please replace the blue text with your own. The paper has three main sections: (1) Overview; (2) Availability; (3) Reuse potential. \\

Please submit the completed paper to: editor.jors@ubiquitypress.com

\rule{\textwidth}{1pt}

\section*{(1) Overview}

\vspace{0.5cm}

\section*{Title}

PyPSA: Python for Power System Analysis

\section*{Paper Authors}

\begin{enumerate}
\item Brown, Tom
\item Hörsch, Jonas
\item Schlachtberger, David
\end{enumerate}

\section*{Paper Author Roles and Affiliations}

All authors are at the Frankfurt Institute for Advanced Studies,
Ruth-Moufang-Straße 1, 60438 Frankfurt am Main, Germany.

Author roles:
\begin{enumerate}
\item Tom wrote most of the paper and is a main developer of PyPSA.
\item Jonas made contributions to the text, provided the performance
  tests and is a main developer of PyPSA.
\item David made contributions to the text and is a main developer of
  PyPSA.
\end{enumerate}

\section*{Abstract}

\else

%

\title{PyPSA: Python for Power System Analysis}

\author{\IEEEauthorblockN{Tom Brown, Jonas Hörsch, David Schlachtberger}
  \IEEEauthorblockA{Frankfurt Institute for Advanced Studies, Ruth-Moufang-Straße 1, 60438 Frankfurt am Main, Germany \\
Email: \href{mailto:brown@fias.uni-frankfurt.de}{brown@fias.uni-frankfurt.de}}
}


%


\maketitle

\begin{abstract}


\fi

  Python for Power System Analysis (PyPSA) is a free software toolbox
  for simulating and optimising modern electrical power systems over multiple
  periods.  PyPSA includes models for conventional generators with
  unit commitment, variable renewable generation, storage units,
  coupling to other energy sectors, and mixed alternating and direct
  current networks. It is designed to be easily extensible and to
  scale well with large networks and long time series. In this paper
  the basic functionality of PyPSA is described, including the
  formulation of the full power flow equations and the multi-period
  optimisation of operation and investment with linear power flow
  equations. PyPSA is positioned in the existing free software
  landscape as a bridge between traditional power flow analysis tools
  for steady-state analysis and full multi-period energy system
  models.  The functionality is demonstrated on two open datasets of
  the transmission system in Germany (based on SciGRID) and Europe
  (based on GridKit).

\ifjors

\section*{Keywords}

Power system simulations; energy system simulations; Load flow
calculations; optimal power flow; security-constrained optimal power
flow; unit commitment; renewable energy.

\else

\end{abstract}
\IEEEpeerreviewmaketitle

\fi

\ifjors
\else
\section{Introduction}
\fi

Power system tools model the interactions between the electrical grid
and the consumers and generators which use the grid. The importance of
software modelling of the grid has risen in recent years given the
increase in distributed and fluctuating wind and solar generation, and
the increasing electrification of all energy demand. On the generation
side, variable renewable generation causes loading in parts of the
grid where it was never expected, and introduces new stochastic
influences on the flow patterns. On the demand side, the need to
decarbonise the transport and heating sectors is leading to the
electrification of these sectors and hence higher electrical demand,
replacing internal combustion engines with electric motors in the
transport sector, and replacing fossil fuel boilers with heat pumps,
resistive heaters and cogeneration for low-temperature space and water
heating. In addition, the increasing deployment of storage
technologies introduces many network users which are both consumers and
generators of energy.

The increasing complexity of the electricity system requires new tools
for power system modelling. Many of the tools currently used for power
system modelling were written in the era before widespread integration
of renewable energy and the electrification of transport and
heating. They therefore typically focus on network flows in single
time periods. Examples of such tools include commercial products like
DIgSILENT PowerFactory \cite{PowerFactory}, NEPLAN \cite{NEPLAN},
PowerWorld \cite{PowerWorld}, PSS/E \cite{PSSE} and PSS/SINCAL
\cite{SINCAL}, and open tools such as MATPOWER \cite{MATPOWER}, PSAT
\cite{PSAT}, PYPOWER \cite{PYPOWER} and pandapower \cite{pandapower}
(see \cite{openelectrical} for a full list of power system analysis
tools).

The consideration of multiple time periods is important on the
operational side for unit commitment of conventional generators and
the optimisation of storage and demand side management, and on the
investment side for optimising infrastructure capacities over
representative load and weather situations. Several tools have subsets
of these capabilities, such as calliope \cite{Pfenninger20171},
minpower \cite{minpower}, MOST \cite{MOST}, oemof \cite{oemof},
OSeMOSYS \cite{Howells20115850}, PLEXOS \cite{PLEXOS}, PowerGAMA
\cite{doi:10.1063/1.4962415}, PRIMES \cite{Primes}, TIMES \cite{Times}
and urbs \cite{johannes_dorfner_2016_60484}, but their representations
of electrical grids are often simplified.


Python for Power System Analysis (PyPSA), the tool
presented in this paper, was developed at the Frankfurt Institute for
Advanced Studies to bridge the gap between power system analysis
software and general energy system modelling tools. PyPSA can model
the operation and optimal investment of the energy system over
multiple periods. It has models for the unit commitment of
conventional generators, time-varying renewable generators,
storage units, all combinations of direct and alternating current
electricity networks, and the coupling of electricity to other energy
sectors, such as gas, heating and transport. It can perform full load flow calculations and linearised
optimal load flow, including under consideration of security
constraints. It was written from the start with variable renewables,
storage and sector-coupling in mind, so that it performs well with
large networks and long time series.

Given the complexity of power system tools and the different needs of
different users, it is crucial that such tools are both transparent in
what they do and easily extendable by the user. To this end, PyPSA was
released as free software under the GNU General Public Licence Version 3 (GPLv3) \cite{gplv3}.
This means that the user is free to
inspect, use and modify the code, provided that if they redistribute the software, they also provide the source code.  Free software and open data also
guarantee that research results can be reproduced by any third party,
which is important given the large investment decisions that will need
to be made on the basis of energy system modelling to reduce
greenhouse gas emissions and combat global warming \cite{PFENNINGER2017211,PfenningerNature}.

PyPSA is available online in the Python Package Index (PyPI), on
GitHub \cite{PyPSA-github} and is archived on Zenodo
\cite{PyPSA-zenodo}.  Documentation and examples are available on
PyPSA's website \cite{PyPSA-website}. PyPSA is already used by more
than a dozen research institutes and companies worldwide, 70 people
are registered on the forum \cite{PyPSA-forum} and the website
\cite{PyPSA-website} has been visited by people from over 160
countries. As of October 2017 it has been used in six research papers
\cite{GorensteinDedecca2017805,Brown2016,2017arXiv170401881H,Schlachtberger2017,Hoersch2017,Groissbock2017}.
Users have already extended PyPSA for integer transmission expansion
\cite{GorensteinDedecca2017805,PyPSA-MILP} and in the grid planning
tool open\_eGo \cite{openego}.

This paper describes version 0.11.0 of PyPSA \cite{PyPSA-zenodo}.
In Section \ref{sec:func} the mathematical functionality of PyPSA is
described, while in Section \ref{sec:software} the focus shifts to the
implementation in software. Quality control is discussed in Section \ref{sec:quality};
the computational performance of PyPSA is described in Section \ref{sec:performance}; and then its functionality is compared with other software in Section \ref{sec:comparison}.
Several example applications are given in
\ref{sec:examples} before conclusions are drawn in \ref{sec:conclusions}.

\ifjors
\else
\section{Functionality}\label{sec:func}
\fi

In this section the basic components, power flow, linear optimal power
flow, energy system optimisation, unit commitment, contingency
modelling and other functionality of PyPSA are described. The
definitions of the main variables used in this section can be found in
Table \ref{tab:variables}, along with units where applicable.

\ra{1.05}

\begin{table}[!t]
	\caption{Nomenclature}
	\label{tab:variables}
	\centering
        \ifjors
        \footnotesize
	\begin{tabular}{@{}llp{10cm}@{}}
        \else
        \begin{tabular}{@{}llp{4.5cm}@{}}
        \fi
\toprule
Variable & Units & Definition \\
\midrule
$n,m$ & & Bus labels \\
$r$ & & Generator energy carrier labels (e.g. wind, solar, gas, etc.) \\
$s$ & & Storage energy carrier labels (e.g. battery, hydrogen, etc.) \\
$k,\ell$ & & Branch labels \\
$c$ & & Cycle labels \\
$t$ & & Snapshot / time point labels \\
$e_{r/s}$ & tCO$_2$eq/MWh\th & CO$_2$-equivalent emissions of energy carrier $r$ or $s$ \\
$w_t$ & h & Weighting of snapshot in objective function \\
$g_{n,r,t}$ & MW & Dispatch of generator at bus $n$ with carrier $r$  at time $t$\\
$G_{n,r}$ & MW & Power capacity of generator $n,r$ \\
$\bar{g}_{n,r,t}$ & MW/MW & Power availability per unit of generator capacity \\
$\eta_{n,r}$ & MW\el/MW\th & Efficiency of generator \\
$u_{n,r,t}$ & & On/off binary status for generator unit commitment\\
$T^{\textrm{min\_down}}_{n,r}$ & h &Generator minimum down time\\
$T^{\textrm{min\_up}}_{n,r}$ & h &Generator minimum up time\\
$ru_{n,r}$ & (MW/MW)/h & Generator ramp up limit per unit of capacity \\
$rd_{n,r}$ & (MW/MW)/h & Generator ramp down limit per unit of capacity \\
$c_{n,r}$ & \euro/MW & Generator capital (fixed) cost \\
$o_{n,r}$ & \euro/MWh & Generator operating (variable) cost \\
$suc_{n,r(,t)}$ & \euro & Generator start up cost (in time $t$) \\
$sdc_{n,r(,t)}$ & \euro & Generator shut down cost (in time $t$) \\
$h_{n,s,t}$ & MW & Dispatch of storage at bus $n$ with carrier $s$ at time $t$ \\
$H_{n,s}$ & MW & Power capacity of storage $n,s$ \\
$e_{n,s,t}$ & MWh & Storage state of charge (energy level) \\
$E_{n,s}$ & MWh & Storage energy capacity \\
$c_{n,s}$ & \euro/MW & Storage power capacity cost \\
$\hat c_{n,s}$ & \euro/MWh & Storage energy capacity cost \\
$o_{n,s}$ & \euro/MWh & Storage dispatch cost \\
$d_{n,t}$ & MW & Electrical load at bus $n$ at time $t$ \\
$\lambda_{n,t}$ & \euro/MWh & Marginal price at bus $n$ at time $t$ \\
$V_n$ & kV & Complex voltage at bus $n$ \\
$\theta_n$ & rad & Voltage angle at bus $n$ \\
$I_n$ & kA & Complex current at bus $n$ \\
$P_n$ & MW & Total active power injection at bus $n$ \\
$Q_n$ & MVAr & Total reactive power injection at bus $n$ \\
$S_n$ & MVA & Total apparent power injection at bus $n$ \\
$f_{\ell,t}$ & MW & Branch active power flow \\
$F_\ell$ & MW & Branch active power rating\\
$c_\ell$ & \euro/MW & Branch capital cost \\
$x_\ell$ & $\Omega$ & Branch series reactance \\
$r_\ell$ & $\Omega$ & Branch series resistance \\
$z_\ell$ & $\Omega$ & Branch series impedance \\
$y_\ell$ & S & Branch shunt admittance \\
$\tau_\ell$ &  & Transformer tap ratio \\
$\theta^{\textrm{shift}}_\ell$ & rad & Transformer phase shift \\
$\eta_{\ell,t}$ & MW/MW & Efficiency loss of a link \\
$K_{n\ell}$ & & $N \times L$ incidence matrix \\
$C_{\ell c}$ & & $L \times (L-N+1)$ cycle matrix \\
$Y_{nm}$ & S & Bus admittance matrix \\
$B_{\ell k}$ & S & Diagonal $L\times L$ matrix of branch susceptances \\
$BODF_{\ell k}$ & & Branch Outage Distribution Factor \\
\bottomrule
	\end{tabular}
\end{table}

\ifjors
\else
\subsection{Components}
\fi

PyPSA's representation of the power system is built by connecting the
components listed in Table \ref{tab:components}.

\ra{1.05}
\begin{table}[t]
	\caption{PyPSA components}
	\label{tab:components}
	\centering
	\begin{tabular}{@{}lp{6cm}@{}}
          \toprule
Network & Container for all other network components. \\
Bus & Fundamental nodes to which all other components attach. \\
Carrier & Energy carrier (e.g. wind, solar, gas, etc.). \\
Load & A consumer of energy. \\
Generator & Generator whose feed-in can be flexible subject to minimum loading or minimum down and up times, or variable according to a given time series of power availability. \\
Storage Unit & A device which can shift energy from one time to another, subject to efficiency losses. \\
Store & A more fundamental storage object with no restrictions on charging or discharging power. \\
Shunt Impedance & An impedance in shunt to a bus. \\
Line & A branch which connects two buses of the same voltage. \\
Transformer & A branch which connects two buses of different voltages. \\
Link & A branch with a controllable power flow between two buses. \\
\bottomrule
	\end{tabular}
\end{table}

Buses are the fundamental nodes to which all other components
attach. Their mathematical role is to enforce energy conservation at
the bus at all times (essentially Kirchhoff's Current Law).

Loads, generators, storage units, stores and shunt impedances attach
to a single bus and determine the power balance at the bus. Loads
represent a fixed power demand; a generator's dispatch can be
optimised within its power availaiblity; stores can shift power from
one time to another with a standing loss efficiency for energy
leakage; storage units behave like stores, but they can also have
efficiency losses and power limits upon charging and discharging;
finally shunt impedances have a voltage-dependent power consumption.

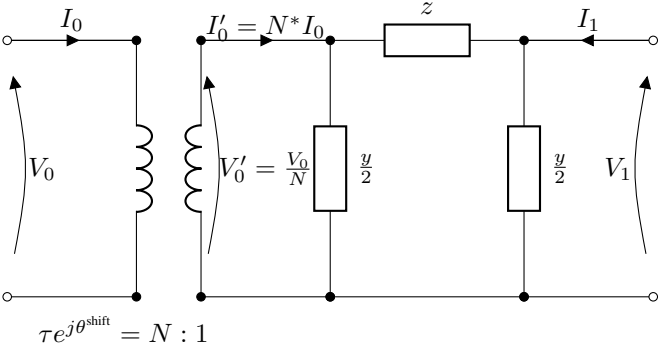
\begin{figure}[t]

\ifjors
\centering
\begin{circuitikz}[scale=1]
\else
\begin{circuitikz}[scale=0.85]
\fi

  \draw (-3,0)
  to [open, v>=$V_0$] (-3,4)
  to [short,i^=$I_0$,o-*] (-1,4) ;

  \draw (-1,0) to [short,*-o] (-3,0);

  \draw (-1,4 ) to [american inductor] (-1,0) ;

  \draw (0,0 ) to [american inductor] (0,4) ;

  \draw (-1.2,-0.5) node{$\tau e^{j\theta^{\textrm{shift}}} = N : 1$};

  \draw (1,2) node{$V_0' =  \frac{V_0}{N}$};

  \draw (1,4.2) node{$I_0' = N^*I_0$};

  \draw
  (0,0)
  to [open, v>=$$] (0,4)
  to [short,i^=$$,*-*] (2,4)
  to [R=$\frac{y}{2}$, -*] (2,0)
  to [short,*-*] (0,0)

  (2,4)
  to [R=$z$] (5,4)
  to [R=$\frac{y}{2}$, -*] (5,0)
  to [short] (2,0)

  (5,0)
  to [short,*-o] (7,0)
  to [open, v^>=$V_1$] (7,4)
  to [short,i_=$I_1$,o-*] (5,4)
  ;
\end{circuitikz}
  \caption{Electrical property definitions for passive branches (lines and transformers).}
\label{fig:equiv}
\end{figure}

Lines and transformers connect two buses with a
given impedance. Power flows through lines and transformers according
to the power imbalances at the buses and the impedances in the
network.
Lines and transformers are referred to collectively as `passive branches'
to distinguish them from controllable link branches.
The impedances of the passive branches are modelled
internally using the equivalent PI model. The relation between the
series impedance $z = r + jx$, the shunt admittance $y = g + jb$, the
transformer tap ratio $\tau$, the transformer phase shift $\theta^{\textrm{shift}}$,
and the complex currents $I_0,I_1$ and complex voltages $V_0,V_1$ at the buses
labelled 0 and 1 is given by
\begin{equation}
  \left( \begin{array}{c}
    I_0 \\ I_1
  \end{array}
  \right) =   \left( \begin{array}{cc}  \frac{1}{z} + \frac{y}{2} &      -\frac{1}{z}\frac{1}{\tau e^{-j\theta^{\textrm{shift}}}}  \\
   -\frac{1}{z}\frac{1}{\tau e^{j\theta^{\textrm{shift}}}} & \left(\frac{1}{z} + \frac{y}{2} \right) \frac{1}{\tau^2}
   \end{array}
   \right)  \left( \begin{array}{c}
    V_0 \\ V_1
  \end{array}
    \right) \label{eq:branchY}
\end{equation}
(For lines, for which neither the tap ratio or the phase shift are
relevant, set $\tau = 1$ and $\theta^{\textrm{shift}} = 0$ in this equation.)
The equivalent circuit is shown in Figure \ref{fig:equiv}.  This
circuit is for the case where the tap-changer is on the primary side;
a similar equation and figure for the case where the tap-changer is on
the secondary side is given in the documentation \cite{PyPSA-website}. The
line model defaults to the PI model, while the transformer model
defaults to the more accurate T model, which is converted to the PI
model using the standard delta-wye transformation. For convenience
standard types for lines and transformers in networks at 50~Hz are provided following
the conversion formula from nameplate parameters to impedances and the typical parameters
provided in pandapower \cite{pandapower}, so that the user does not have to
input the impedances manually. The typical parameters in pandapower are based on \cite{heuck,ooe,oswald}.

Links connect two buses with a controllable active power dispatch that can be
set by the user or optimised by PyPSA. Links can be used to represent
point-to-point high voltage direct current (HVDC) lines, import-export capacities in transport models such as Net-Transfer-Capacity (NTC) models, or general
energy conversion processes with a given efficiency, such as resistive heaters or heat pumps (from
electricity to heat) or gas boilers (from gas to heat). Their
efficiency can also be time-varying (e.g. to represent the ambient
temperature dependence of a heat pump's coefficient of performance).
Networks of links implement Kirchoff's Current Law (energy conservation at each bus), but not Kirchoff's Voltage Law, which is obeyed by networks of passive branches.

A generator can also be represented in terms of more basic components: a bus is added for the fuel source with a store to represent the amount of fuel available. It is then connected to the electricity bus with a link to represent the energy conversion loss. Similarly a storage unit can be represented with an additional bus for the storage medium with a store attached, and then two links connected to the electricity bus to represent charging and discharging.

Energy enters the model in generators; in storage units or stores with
higher energy levels before than after the simulation; and in any
components with efficiency greater than 1 (such as heat pumps).
Energy leaves the model in loads; in storage units or stores with
higher energy levels after than before the simulation; and in lines,
links or storage units with efficiency less than 1.

\ifjors
\else
\subsection{Power flow without optimisation}
\fi

In a power flow calculation, the user specifies the power dispatch of
all dispatchable components (loads, generators, storage units, stores and links)
and then PyPSA computes the resulting voltages in the network and
hence the power flows in passive branches (lines and transformers)
based on their impedances.

\ifjors
\else
\subsubsection{Power flow equations for AC networks}
\fi

A power flow calculation for an alternating current (AC) network
ensures that for all buses labelled by $n$ we have
\begin{equation}
  S_n = V_nI_n^* = \sum_{m} V_n Y_{nm}^* V_m^* \label{eq:pf}
\end{equation}
where $S_n = P_n + jQ_n$ is the apparent power injected at the bus,
$I_n$ is the complex current and $V_n = |V_n|e^{j\theta_n}$ is the
complex voltage, whose rotating angle is measured relative to a chosen
slack bus. $Y_{nm}$ is the bus admittance matrix, which is constructed
for all buses based on the contributions from the individual branch
admittance matrices from equation \eqref{eq:branchY} and any shunt impedances
at the nodes, following the example of MATPOWER \cite{MATPOWER}.

The inputs and outputs for the buses are given as follows:
\begin{itemize}
  \item For the chosen slack bus $n=0$, it is assumed that the voltage magnitude $|V_0|$ and  the voltage angle $\theta_0$ are given. PyPSA must find the powers $P_0$ and $Q_0$.
  \item For $PQ$ buses, $P_n$ and $Q_n$ are given; $|V_n|$ and $\theta_n$ are to be found.
  \item For $PV$ buses, $P_n$ and $|V_n|$ are given; $Q_n$ and $\theta_n$ are to be found.
\end{itemize}

The non-linear equation system \eqref{eq:pf} is then solved using the
Newton-Raphson algorithm \cite{Grai94} and, by default, an initial `flat' guess of $\theta_n =
\theta_0$ and $|V_n| = 1$ (per unit). The initial guess can also be specified (`seeded') by the user, using for example the linearised power flow solution.

\ifjors
\else
\subsubsection{Power flow equations for DC networks}
\fi

A power flow calculation for a direct current (DC) network
ensures that for all buses labelled by $n$ we have
\begin{equation}
  P_n = V_nI_n = \sum_{m} V_n G_{nm} V_m \label{eq:dcpf}
\end{equation}
where $P_n$ is the active power injected at the bus and the voltage,
current and the conductance matrix $G_{ij}$ are now all real
quantities. This non-linear equation is also solved with the
Newton-Raphson algorithm.

\ifjors
\else
\subsubsection{Linearised power flow equations for AC networks}
\fi

In some circumstances a linearisation of the AC power flow equations
\eqref{eq:pf} can provide a good approximation to the full non-linear
solution \cite{Purc05,Stot09}.  The linearisation is restricted to calculating active power flows based on
voltage angle differences and branch series reactances. It assumes
that reactive power flow decouples from active power flow, that there
are no voltage magnitude variations, voltage angles differences across
branches are small enough that $\sin \theta \sim \theta$ and branch
resistances are negligible compared to branch reactances. This makes
it suitable for short overhead transmission lines close to their
natural loading.

In this case it can be shown \cite{MATPOWER} that the voltage angles are related to the active power injections by a matrix
\begin{equation}
   P_n = \sum_m (KBK^T)_{nm} \theta_m - \sum_\ell K_{n\ell} b_\ell \theta_\ell^{\textrm{shift}}
\end{equation}
where $K$ is the incidence matrix of the network, $B$ is the diagonal
matrix of inverse branch series reactances $x_\ell$ multiplied by the
tap ratio $\tau_\ell$, i.e. $B_{\ell\ell} = b_\ell =
\frac{1}{x_\ell\tau_\ell}$, and $\theta_\ell^{\textrm{shift}}$ is the
phase shift for a transformer. The matrix $KBK^T$ is singular with a
single zero eigenvalue for a connected network and can be inverted by first deleting the row and
column corresponding to the slack bus.

\ifjors
\else
\subsubsection{Linearised power flow equations for DC networks}
\fi

For DC networks the equation \eqref{eq:dcpf} is linearised by positing
$V_n = 1 + \delta V_n$ and assuming that $\delta V_n$ is small. The
resulting equations mirror the linearised AC approximation with the
substitutions $\theta_n \to \delta V_n$ and $x_\ell \to r_\ell$.

\ifjors
\else
\subsection{Optimisation with linear power flow equations}
\fi

PyPSA is a partial equilibrium model that can optimise both short-term operation and long-term investment in the  energy system as a linear problem using the linear power flow equations.

PyPSA minimises total system costs, which include
the variable and fixed costs of generation, storage and transmission,
given technical and physical constraints. The objective function is given by

\ifjors
\begin{eqnarray}
\else
\begin{IEEEeqnarray}{rCl}
\fi
&& \min_{\substack{F_\ell,G_{n,r},H_{n,s}, E_{n,s} \\ f_{\ell,t},g_{n,r,t},h_{n,s,t},suc_{n,r,t},sdc_{n,r,t}}} \left[ \sum_{\ell} c_{\ell} \cdot F_{\ell}  + \sum_{n,r} c_{n,r} \cdot G_{n,r} \right.\nonumber  \\
  && + \sum_{n,r,t} \left( w_t \cdot o_{n,r} \cdot g_{n,r,t} + suc_{n,r,t} + sdc_{n,r,t} \right)  \\
&& \left. + \sum_{n,s} c_{n,s} \cdot H_{n,s} + \sum_{n,s} \hat{c}_{n,s} \cdot E_{n,s} + \sum_{n,r,t}  w_t \cdot o_{n,s} \cdot [h_{n,s,t}]^+   \right] \nonumber
\ifjors
\end{eqnarray}
\else
\end{IEEEeqnarray}
\fi
It consists of the branch capacities $F_\ell$ for each branch
$\ell$ and their annuitised fixed costs per capacity $c_\ell$, the generator
capacities $G_{n,r}$ at each bus $n$ for technology $r$ and their
annuitised fixed costs per capacity $c_{n,r}$, the dispatch $g_{n,r,t}$  of the unit at time $t$ and the associated variable costs $o_{n,r}$,
the start up and shut down costs $suc_{n,r,t}$ and $sdc_{n,r,t}$ when unit commitment is activated,
the storage unit power capacities $H_{n,s}$ and store energy capacities $E_{n,s}$ at each bus $n$ for storage technology $s$ and their associated fixed costs $c_{n,s}$ and $\hat{c}_{n,s}$, and finally
the positive part of the storage dispatch $[h_{n,s,t}]^+$ and the associated variable costs $o_{n,s}$.
The branch flows $f_{\ell,t}$ are optimisation variables but do not appear in the objective function.
The optimisation is run over multiple time periods $t$ representing different weather and demand conditions. Each period  can have a weighting $w_t$; the investment costs must then be annuitised for the total period $\sum_t w_t$ (typically a full year).

The dispatch of generators $g_{n,r,t}$ is constrained by their
capacities $G_{n,r}$ and time-dependent availabilities
$\tilde{g}_{n,r,t}$ and $\bar{g}_{n,r,t}$, which are given per unit of
the capacities $G_{n,r}$:
\begin{equation}
 \tilde{g}_{n,r,t} \cdot G_{n,r} \leq  g_{n,r,t} \leq \bar{g}_{n,r,t} \cdot G_{n,r} \hspace{1cm} \forall\, n,r,t \label{eq:gen}
\end{equation}
For conventional generators the availabilities are usually constant;
a fully flexible generator would have $\tilde{g}_{n,r,t} = 0$ and
$\bar{g}_{n,r,t} = 1$. For variable renewable generators such as wind and solar, $\bar{g}_{n,r,t}$ represents the weather-dependent power availability,
while curtailment may also be limited by introducing a lower bound on the dispatch $\tilde{g}_{n,r,t}$.

The dispatch can also be limited by ramp rate constraints $ru_{n,r}$ and $rd_{n,r}$ per unit of
the generator nominal power:
\begin{equation}
   -rd_{n,r}\cdot G_{n,r} \leq (g_{n,r,t} - g_{n,r,t-1}) \leq ru_{n,r}\cdot G_{n,r} \hspace{.5cm} \forall\, n,r,t>0\label{eq:ramp}
\end{equation}
Unit commitment for conventional generators is described in Section \ref{sec:unit}.

The power capacity $G_{n,r}$ can also be optimised within minimum $\tilde{G}_{n,r}$ and maximum $\bar{G}_{n,r}$ installable potentials:
\begin{equation}
 \tilde{G}_{n,r} \leq G_{n,r}\leq \bar{G}_{n,r} \hspace{1cm} \forall\, n,r
\end{equation}

The dispatch of storage units $h_{n,s,t}$, whose energy carriers are labelled by
$s$, is constrained by a similar equation to that for generators in
equation \eqref{eq:gen}:
\begin{equation}
 \tilde{h}_{n,s,t} \cdot H_{n,s} \leq  h_{n,s,t} \leq \bar{h}_{n,s,t} \cdot H_{n,s} \hspace{1cm} \forall\, n,s,t
\end{equation}
except $\tilde{h}_{n,s,t}$ is now negative, since the dispatch of
storage units can be both positive when discharging into the grid and
negative when absorbing power from the grid. The power capacity $H_{n,s}$ can also be optimised within installable potentials.

The energy levels $e_{n,s,t}$ of all storage units have to be consistent between all hours and are limited by the storage energy capacity $E_{n,s}$
\ifjors
\begin{eqnarray}
\else
\begin{IEEEeqnarray}{rCl}
\fi
  e_{n,s,t} & = & \eta_{n,s,0}^{w_t} e_{n,s,t-1}\nonumber \\
  & & + \eta_{n,s,+}\cdot w_t \left[h_{n,s,t}\right]^+  -  \eta_{n,s,-}^{-1}\cdot w_t\left[h_{n,s,t}\right]^-\nonumber \\
  & & + w_t\cdot h_{n,s,t,\textrm{inflow}} - w_t\cdot h_{n,s,t,\textrm{spillage}} \nonumber \\
 \tilde{e}_{n,s,t}\cdot E_{n,s} & \leq &  e_{n,s,t} \leq \bar{e}_{n,s,t}\cdot E_{n,s}   \hspace{1cm} \forall\, n,s,t
\ifjors
\end{eqnarray}
\else
\end{IEEEeqnarray}
\fi
Positive and negative parts of a value are denoted as
$[\cdot]^+= \max(\cdot,0)$, $[\cdot]^{-} = -\min(\cdot,0)$.  The
storage units can have a standing loss (self-discharging leakage rate) $\eta_{n,s,0}$, a charging efficiency
$\eta_{n,s,+}$, a discharging efficiency $\eta_{n,s,-}$, inflow (e.g. river inflow
in a reservoir) and spillage. The initial energy level can be set by the user, or it is assumed to be
cyclic, i.e. $e_{n,s,t=0} = e_{n,s,t=T}$.

The store component is a more basic version of the storage unit: its charging and discharging power cannot be limited and there are no charging and discharging efficiencies $\eta_{n,s,+},\eta_{n,s,-}$. The energy levels of the store can also be restricted by time series $\tilde{e}_{n,s,t}, \bar{e}_{n,s,t}$ given per unit of the energy capacity $E_{n,s}$; this allows the demand-side management model of \cite{2014arXiv1401.4121K} to be implemented in PyPSA. The energy capacity $E_{n,s}$ can also be optimised within installable potentials.

Global constraints related to primary energy consumption, such as emission limits, can also be implemented. For example, CO${}_2$ emissions can be limited by a cap $\textrm{CAP}_{CO2}$, implemented using the
specific emissions $e_{r}$ in CO${}_2$-tonne-per-MWh\th of the fuel $r$
and the efficiency $\eta_{n,r}$ of the generator:
\begin{equation}
  \sum_{n,r,t} \frac{1}{\eta_{n,r}} w_t \cdot g_{n,r,t}\cdot e_{r} \leq  \textrm{CAP}_{CO2} \quad \leftrightarrow \quad \mu_{CO2} \label{eq:co2cap}
\end{equation}
$\mu_{CO2}$ is the shadow price of this constraint.

The (inelastic) electricity demand $d_{n,t}$ at each bus $n$ must be met at each time $t$ by either local generators and storage or by the flows $f_{\ell,t}$ from the branches $\ell$
\ifjors
\begin{eqnarray}
\else
\begin{IEEEeqnarray}{rCl}
\fi
\sum_{r} g_{n,r,t} + \sum_{s} h_{n,s,t} +  \sum_{\ell} \alpha_{\ell, n,t}\cdot f_{\ell,t}  & = & d_{n,t}   \label{eq:balance}\\
& \leftrightarrow &   w_t\cdot \lambda_{n,t} \hspace{0.3cm} \forall\, n,t \nonumber
\ifjors
\end{eqnarray}
\else
\end{IEEEeqnarray}
\fi
where $\alpha_{\ell,n,t}=-1$ if $\ell$ starts at $n$, $\alpha_{\ell,n,t}=1$ if $\ell$ is a line or transformer and ends at $n$, and $\alpha_{\ell,n,t} = \eta_{\ell,t}$ if $\ell$ is a link and ends at $n$ (note that for lines and transformers, $\alpha_{\ell,n,t}$ is the incidence matrix of the network, $\alpha_{\ell,n,t} = K_{n\ell}$). $\eta_{\ell,t}$ represents an efficiency loss for a link (it can be time-dependent for efficiency that, for example, depends on the outside temperature, like for a heat pump). $\lambda_{n,t}$ is the marginal price at the bus. This
equation implements Kirchhoff's Current Law (KCL), which guarantees energy conservation at each node.

The flows in all passive branches are constrained by their capacities $F_{\ell}$
\begin{equation}
  |f_{\ell,t}| \leq F_{\ell} \hspace{1cm} \forall\,\ell,t
\end{equation}
For links, the flows can be more finely controlled with time-dependent per unit availabilities
$\tilde{f}_{\ell,t}, \bar{f}_{\ell,t}$
\begin{equation}
 \tilde{f}_{\ell,t} \cdot F_{\ell}  \leq f_{\ell,t} \leq \bar{f}_{\ell,t} \cdot F_{\ell} \hspace{1cm} \forall\,\ell,t
\end{equation}
which allows, for example, time-dependent demand-side management schemes to be modelled \cite{2014arXiv1401.4121K}. For both passive branches and links, the upper and lower limits are associated with KKT
multipliers $\bar{\mu}_{\ell,t}$ and $\ubar{\mu}_{\ell,t}$.

The flows in links are fully controllable.

Power flows in networks of passive branches (lines and transformers)
according to the linear power flow equations. It is assumed that the network is lossless, so that $\eta_{\ell,t} = 1$ for passive branches.
To guarantee the physicality of the network
flows, in addition to KCL, Kirchhoff's Voltage Law (KVL) must be
enforced in each connected network.  KVL states that the voltage
differences around any closed cycle in the network must sum to
zero. If each independent cycle $c$ is expressed as a directed combination
of passive branches $\ell$ by a matrix $C_{\ell c}$ then KVL becomes the constraint
\begin{equation}
  \sum_{\ell} C_{\ell c} \cdot x_{\ell} \cdot f_{\ell,t} = 0 \quad  \hspace{1cm} \forall c,t \label{eq:kvl}
\end{equation}
where $x_\ell$ is the series inductive reactance of branch $\ell$.
In a recent paper it is demonstrated that this formulation of the linear load flow using cycles solves up to 20 times faster than standard formulations using the voltage angles \cite{2017arXiv170401881H}; voltage angle and PTDF formulations are also implemented in PyPSA and deliver identical results.

Since branch capacities $F_{\ell}$ can be continuously expanded to
represent the addition of new circuits to an aggregated transmission corridor $\ell$, the impedances $x_\ell$ of the
branches would also decrease. In principle this would introduce a
bilinear coupling in equation (\ref{eq:kvl}) between the $x_\ell$ and the $f_{\ell,t}$. To keep the optimisation problem linear and therefore computationally fast, $x_\ell$ can be left fixed in each optimisation problem, updated and then the optimisation problem rerun, in up to 5 iterations to ensure convergence, following the methodology of \cite{Hagspiel}. Another author has implemented an integer transmission expansion in PyPSA \cite{PyPSA-MILP} that bypasses the bilinearity with a disjunctive big-$M$ relaxation \cite{Bahiense2001}; this will be incorporated into the main code base of PyPSA soon.

\ifjors
\else
\subsection{Coupling to other energy sectors}\label{sec:coupling}
\fi


\begin{figure}[!t]
  \ifjors
  \centering
  \begin{adjustbox}{scale=1,trim=5 6.8cm 0 0}
  \else
  \begin{adjustbox}{scale=0.60,trim=5 6.8cm 0 0}
  \fi

  \begin{circuitikz}

  \draw (1.5,14.5) to [short,i^=grid connection] (1.5,13);
  \draw [ultra thick] (-5,13) node[anchor=south]{electric bus} -- (6,13);
  \draw(2.5,13) |- +(0,0.5) to [short,i^=$$] +(2,0.5);
  \draw (0,-0.5) ;
  \draw (0.5,13) -- +(0,-0.5) node[sground]{};
  \draw (2.5,12) node[vsourcesinshape, rotate=270](V2){}
  (V2.left) -- +(0,0.6);
  \draw (2.5,11.2) node{generators};
    \node[draw,minimum width=1cm,minimum height=0.6cm,anchor=south west] at (3.4,11.9){storage};
    \draw (4,13) to (4,12.5);

  \draw [ultra thick] (-6,10) node[anchor=south]{transport} -- (-3,10);
  \draw (-5.5,10) -- +(0,-0.5) node[sground]{};
  \draw (-3.5,10) to [short,i_=${}$] (-3.5,13);
  \draw (-3.2,11.5)  node[rotate=90]{discharge};
  \draw (-4.5,13) to [short,i^=${}$] (-4.5,10);
  \draw (-4.2,11.5)  node[rotate=90]{charge};
  \node[draw,minimum width=1cm,minimum height=0.6cm,anchor=south west] at (-4.5,8.9){battery};
  \draw (-4,10) to (-4,9.5);

    \draw [ultra thick] (2,10) -- (6.5,10)  node[anchor=south]{heat};
  \draw (3.5,10) -- +(0,-0.5) node[sground]{};
  \draw (4.5,9.35) to [esource] (4.5,8.5);
  \draw (4.5,10) -- (4.5,9.35);
  \draw (4.5,8.3) node{solar thermal};
  \draw (5,13) to [short,i^=heat pump;] (5,10);
  \draw (6.2,11) node{resistive heater};
  \node[draw,minimum width=1cm,minimum height=0.6cm,anchor=south west] at (5.5,8.9){hot water tank};
  \draw (6,10) to (6,9.5);

  \draw [ultra thick] (-2,10)  -- (0.5,10) node[anchor=south]{hydrogen};
  \draw (-1.5,13) to [short,i_=${}$] (-1.5,10);
    \draw (-1.2,11.5)  node[rotate=90]{electrolysis};
  \draw (-0.5,10) to [short,i^=${}$] (-0.5,13);
  \draw (-0.2,11.5)  node[rotate=90]{fuel cell};
  \draw (-1,10) to (-1,9.5);
  \node[draw,minimum width=1cm,minimum height=0.6cm,anchor=south west] at (-1.5,8.9){store};

  \draw (0,10) to [short,i_=${}$] (0,8);
  \draw [ultra thick] (-1,8) node[anchor=south]{methane} -- (3,8);
  \draw (1.5,8) to [short,i_=${}$] (1.5,13);
  \draw (2.5,8) to [short,i_=${}$] (2.5,10);
  \node[draw,minimum width=1cm,minimum height=0.6cm,anchor=south west] at (0.5,6.9){store};
  \draw (1,8) to (1,7.5);
  \draw (0.3,9)  node[rotate=90]{Sabatier};
  \draw (1.8,9.2)  node[rotate=90]{generator/CHP};
  \draw (2.8,9)  node[rotate=90]{boiler/CHP};

  \end{circuitikz}

\end{adjustbox}
\caption{Example of the coupling in PyPSA between electricity (at top) and other energy sectors: transport, hydrogen, natural gas and heating. There is a bus for each energy carrier, to which different loads, energy sources and converters are attached.}
\label{fig:flow}
\end{figure}
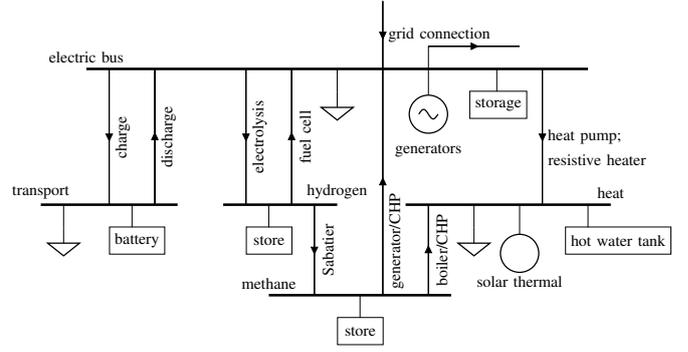

PyPSA can also optimise operation and investment in other energy
sectors, such as natural gas, heating and transport. These sectors can
be modelled using a network of links with efficiencies for energy
conversion losses; an example from a recent paper \cite{Brown2016} is
shown in Figure \ref{fig:flow}. For example, links from electricity to heat buses
can represent resistive heaters and/or heat pumps (the latter can also be
modelled with a time-dependent coefficient of performance, given the
importance of capturing the dependence of heat pump performance on
outside temperature \cite{PETROVIC2016787}). Combined Heat and Power plants (CHPs)
can also be modelled by adding additional constraints for the back
pressure and fuel consumption (see the PyPSA examples \cite{PyPSA-website}).
Depletable resources such as natural gas are modelled with stores.

\ifjors
\else
\subsection{Unit Commitment}\label{sec:unit}
\fi

Unit commitment can be turned on for any generator. This introduces a
times series of new binary status variables $u_{n,r,t} \in \{0,1\}$,
which indicates whether the generator is running (1) or not (0) in
period $t$. The restrictions on generator output now become:
\begin{equation}
 u_{n,r,t}\cdot\tilde{g}_{n,r,t}\cdot G_{n,r} \leq g_{n,r,t} \leq   u_{n,r,t} \cdot\bar{g}_{n,r,t} \cdot G_{n,r} \hspace{.3cm} \forall\, n,r,t
\end{equation}
so that if $ u_{n,r,t} = 0$ then also $g_{n,r,t} = 0$.

If $T^{\textrm{min\_up}}_{n,r}$ is the minimum up time then we have
\begin{equation}
  \sum_{t'=t}^{t+T^\textrm{min\_up}_{n,r}} u_{n,r,t'}\geq T^\textrm{min\_up}_{n,r} (u_{n,r,t} - u_{n,r,t-1})   \hspace{.5cm} \forall\, n,r,t
\end{equation}
(i.e. if the generator has just started up ($u_{n,r,t} - u_{n,r,t-1} = 1$) then it has to run for at least $T^{\textrm{min\_up}}_{n,r}$ periods). Similarly for a minimum down time of $T^{\textrm{min\_down}}_{n,r}$
\begin{equation}
  \sum_{t'=t}^{t+T^\textrm{min\_down}_{n,r}} (1-u_{n,r,t'})\geq T^\textrm{min\_down}_{n,r} (u_{n,r,t-1} - u_{n,r,t})   \hspace{.5cm} \forall\, n,r,t
\end{equation}

For non-zero start up costs $suc_{n,r}$ a new variable $suc_{n,r,t} \geq 0$ is introduced for each time period $t$ and added to the objective function.  The variable satisfies
\begin{equation}
  suc_{n,r,t} \geq suc_{n,r} (u_{n,r,t} - u_{n,r,t-1})   \hspace{.5cm} \forall\, n,r,t
\end{equation}
so that it is only non-zero if $u_{n,r,t} - u_{n,r,t-1} = 1$, i.e. the generator has just started, in which case the inequality is saturated $suc_{n,r,t} = suc_{n,r}$. Similarly for the shut down costs $sdc_{n,r,t} \geq 0$ we have
\begin{equation}
  sdc_{n,r,t} \geq sdc_{n,r} (u_{n,r,t-1} - u_{n,r,t})   \hspace{.5cm} \forall\, n,r,t
\end{equation}
The ramp-rate limits in equation \eqref{eq:ramp} can also be
suplemented by ramping limits at start-up and shut-down.

\ifjors
\else
\subsection{Security-Constrained LOPF}
\fi

PyPSA has functionality to examine the steady state of the power
system after outages of passive branches, based on an analysis of the
linear power flow.

PyPSA calculates the Branch Outage Distribution Factor (BODF) from the
Power Transfer Distribution Factors (PTDF) (see
\cite{WoodWollenberg1996}). The BODF gives the change in linearised power
flow on passive branch $\ell$ given the outage of passive branch $k$
\begin{equation}
  f_{\ell}^{(k)} = f_\ell + BODF_{\ell k} \cdot f_{k}
\end{equation}
Here $f_\ell$ is the flow before the outage and $f_\ell^{(k)}$ is the flow after the outage of branch $k$.

The BODF can then be used in Security-Constrained Linear Optimal Power
Flow (SCLOPF). SCLOPF builds on the LOPF by including additional
constraints that branches may not become overloaded after the outage
of a selection of branches. For each potential outage of a branch $k$, a set of
constraints for all other branches $\ell$ is included, guaranteeing that
they do not become overloaded beyond their capacity $F_\ell$
\begin{equation}
  |f_{\ell,t}^{(k)}| = |f_{\ell,t} + BODF_{\ell k}\cdot f_{k,t}| \leq |F_\ell| \hspace{1cm} \forall \ell, t
\end{equation}

\ifjors
\else
\subsection{Network clustering}
\fi

PyPSA also implements a variety of network clustering algorithms to
reduce the number of buses in a network while preserving important
transmission lines. For example, the $k$-means clustering algorithm
was recently used in \cite{Hoersch2017} to examine the effect of
clustering on investment optimisation results.

\ifjors
\else
\subsection{Planned new features}
\fi

PyPSA is currently in version 0.11.0. PyPSA has been designed to be
modular, so that it is possible to develop the code for many other
types of calculations. Currently features being considered by the
development team include, in rough order of priority:
\begin{itemize}
  \item Integer transmission expansion, following an existing implementation in PyPSA \cite{PyPSA-MILP} using the disjunctive big-$M$ relaxation \cite{Bahiense2001};
  \item Multi-horizon dynamic investment optimisation over several years, following for example the implementation in OSeMOSYS  \cite{Howells20115850};
  \item Transient analysis using the Root-Mean-Square (RMS) values of phasor quantities, following the implementation in PSAT \cite{PSAT};
  \item An implementation of the non-linear power flow solution using analytic continuation in the complex plane \cite{trias2012holomorphic}, following the implementation in GridCal \cite{GridCal};
  \item Short-circuit analysis, following the implementation in pandapower \cite{pandapower};
    \item OPF with the full non-linear network equations, following the implementations in PYPOWER and MATPOWER;
  \item An interactive web-based GUI for analysing and manipulating the network topology.
\end{itemize}

\ifjors
\else
\section{Implementation and architecture}\label{sec:software}
\fi

PyPSA was written in the Python programming language \cite{Python}
because it is widely used in the modelling community, it is easy to
learn and its implementation is also free. It is available for every
major operating system, including GNU/Linux, Mac OSX and
Windows. PyPSA has been tested with versions 2.7 and 3.5 of Python.

PyPSA stores all data about network components in the DataFrame
objects of the Python library pandas \cite{pandas}. This enables easy
and efficient indexing of the data, while mantaining the fast
calculation speeds of the underlying array objects of the Python library
NumPy \cite{NumPy}. For each of the components listed in Table
\ref{tab:components} (except the overall Network container component) there is a
DataFrame listing the static attributes (such as line impedance or
capital cost) and a dictionary of DataFrames containing the
time-varying attributes (such as wind power availability or consumer
demand) that are in addition indexed by the list of snapshots. The
specification of some attributes (such as generator maximum output)
can be either static or time-varying; if the time series is not
specified, then the static value is taken.

All matrix calculations and solutions of linear equation systems are
carried out either with NumPy \cite{NumPy} or, in the case of sparse
matrices, with SciPy \cite{SciPy}. These Python libraries interface with
lower-level programming language libraries to benefit from the speed
of strongly-typed languages.

Optimisation problems are formulated using the Python-based
optimization modeling language Pyomo
\cite{hart2011pyomo,hart2012pyomo}, which is solver agnostic and
intuitive to extend. The use of Pyomo also allows inter-operability
with other energy system frameworks that use Pyomo, such as calliope \cite{Pfenninger20171}, oemof \cite{oemof} and urbs \cite{johannes_dorfner_2016_60484}. In PyPSA lower-level functions in Pyomo have been exploited to improve computational performance.

\begin{figure*}[t]
  \ifjors
\else
  \includegraphics[trim=0 0 0.5cm 0,clip=true,height=4.5cm]{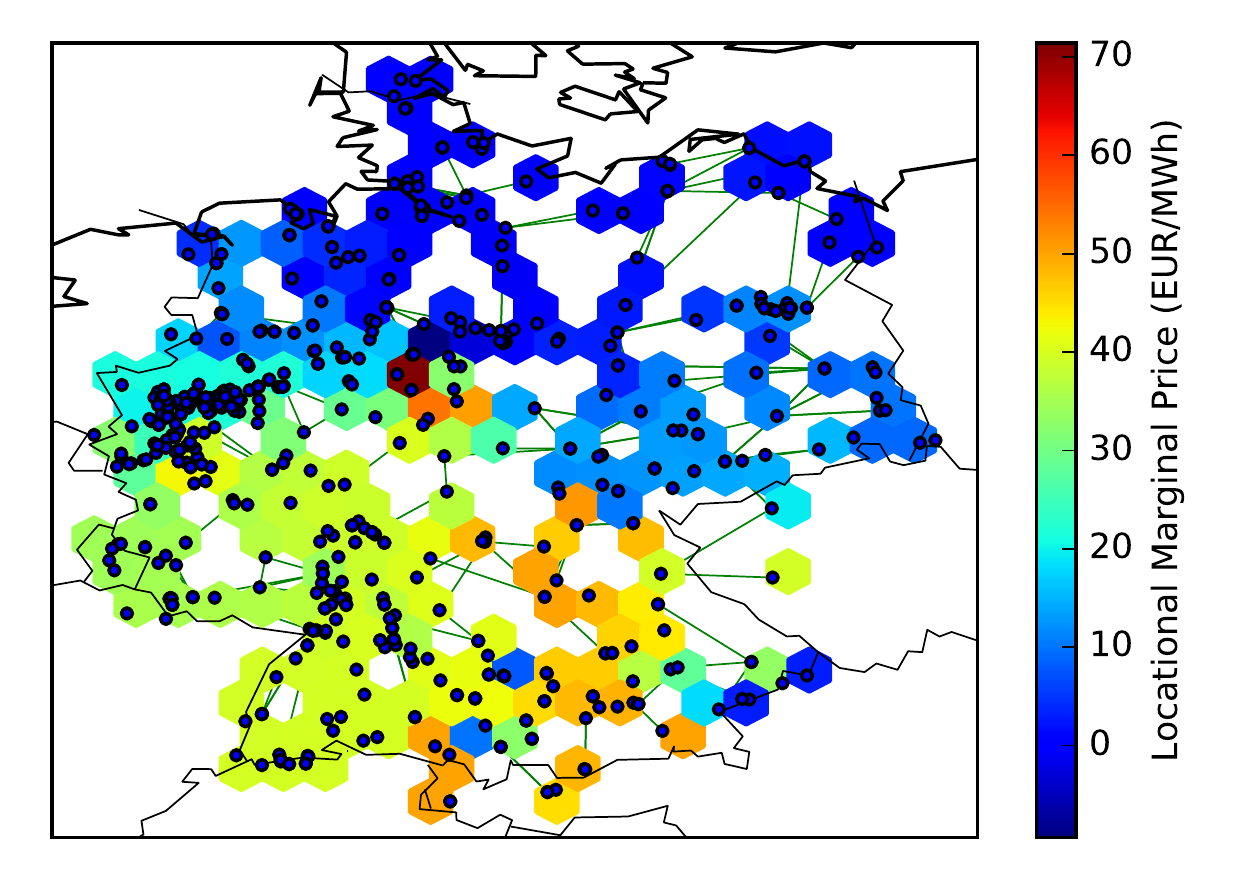}
  \includegraphics[trim=0 0 0.5cm 0,clip=true,height=4.5cm]{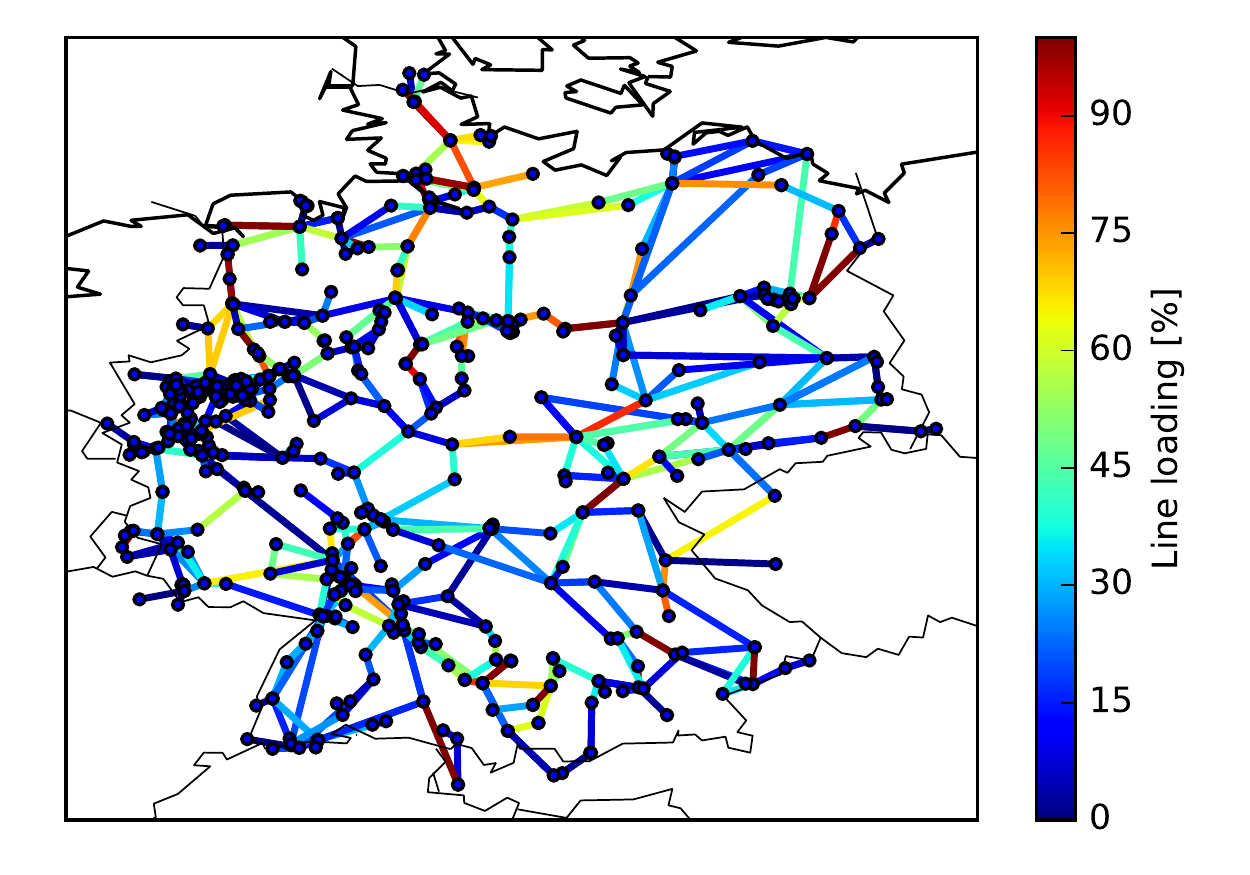}
  \includegraphics[trim=1.5cm 0 1.5cm 0,clip=true,height=4.5cm]{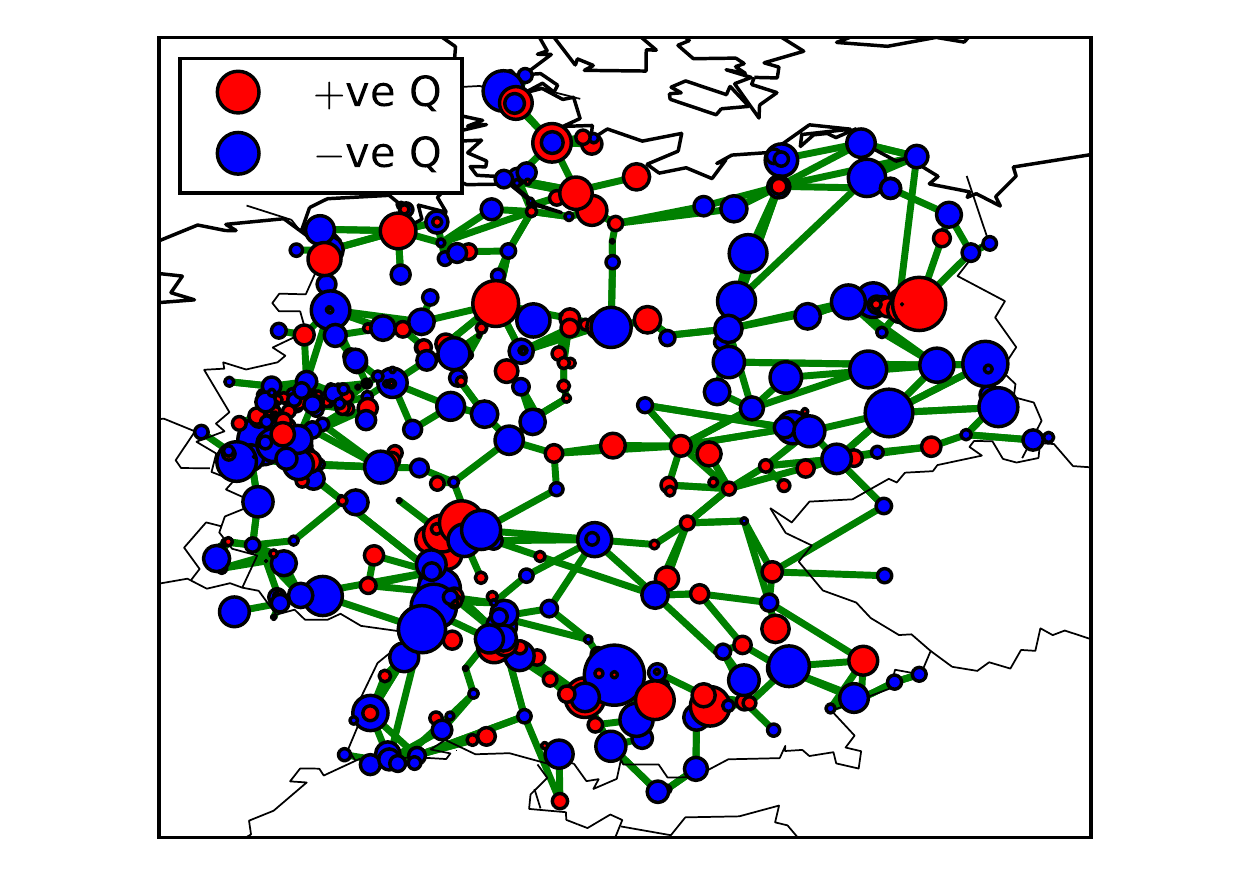}
  \fi
  \caption{\emph{Left:} Locational marginal prices ($\lambda_{n,t}$ from equation \eqref{eq:balance})  for Germany in an hour with high wind and low load; \emph{Middle:} Line loading during this hour: highly loaded lines in the middle of Germany prevent the transport of cheap wind energy to consumers in the South; \emph{Right:} Reactive power feed-in (positive in red, negative in blue) necessary to keep all buses at unit nominal voltage. }
\label{fig:scigrid}
\end{figure*}

\begin{figure}[t]
  \includegraphics[trim=0 3.cm 0 4.5cm,clip=true,width=\linewidth]{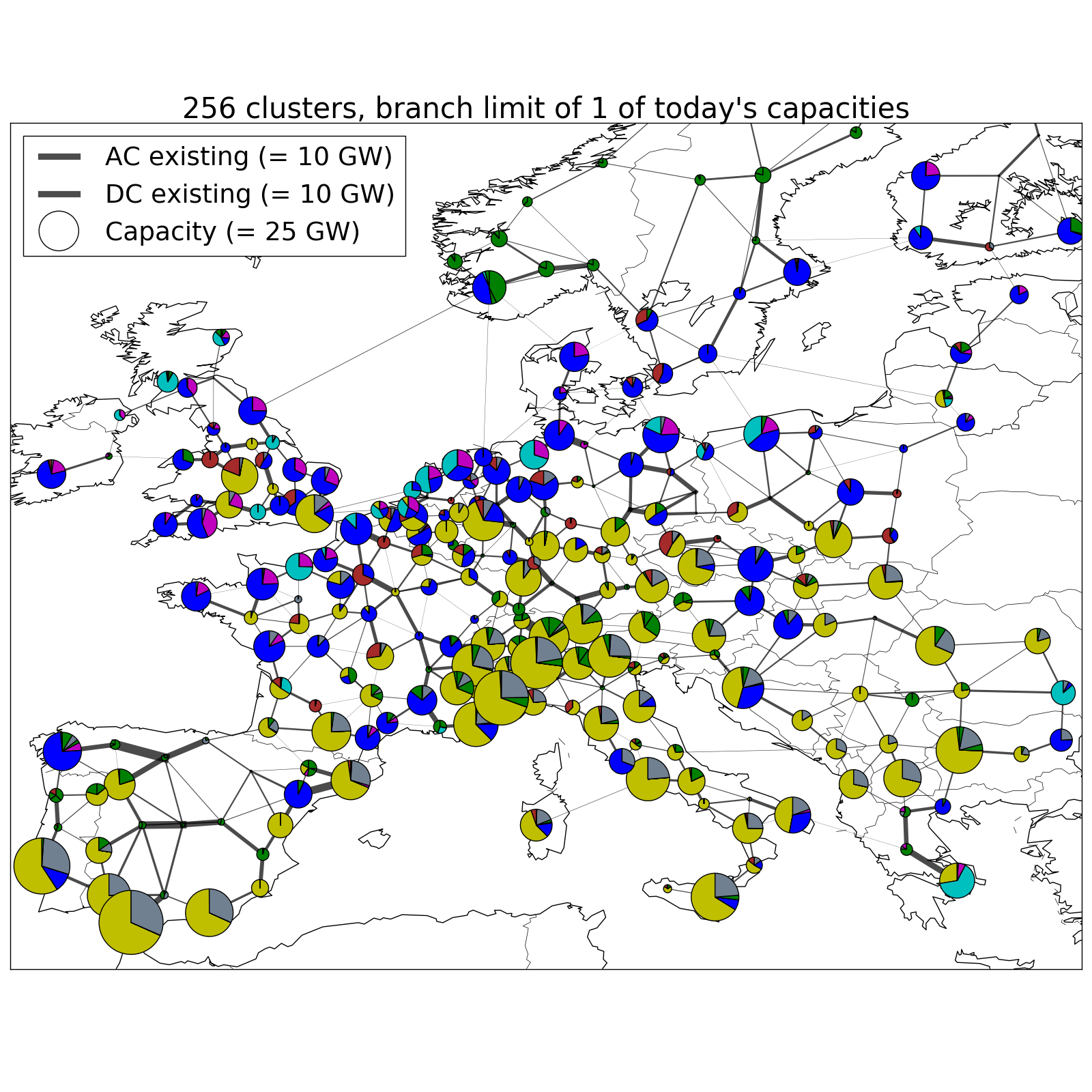}

  \includegraphics[width=\linewidth]{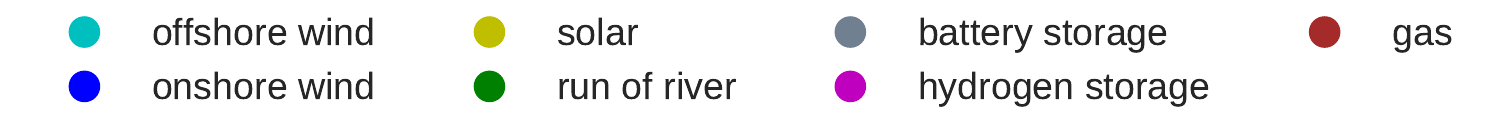}
  \caption{Results of optimisation of generation and storage
  capacities in Europe to reduce CO${}_2$ emissions in the European
  electricity sector by 95\% compared to 1990 levels
  \cite{Hoersch2017}. The grid topology is based on the GridKit network for Europe, clustered from 5000 buses to 256 buses. }
\label{fig:europe}
\end{figure}

PyPSA has no graphical user interface, but integrates closely with the
IPython \cite{IPython} interactive notebooks, where networks and their
properties can be visualised using the Python library Matplotlib
\cite{Matplotlib} (see Figures \ref{fig:scigrid} and \ref{fig:europe})
or the interactive JavaScript-based library plotly \cite{plotly}.


Internally PyPSA converts all power system quantities (voltage, power,
current, impedances) to per unit values. Set points for loads and
generation are stored separately from the power values which are
actually dispatched.

\ifjors
\else
\section{Quality control}\label{sec:quality}
\fi

PyPSA comes with a large test suite that covers all of its major
functionality. Tests are implemented using the Python library pytest
\cite{pytest}. Tests are also included that compare PyPSA's results
with other software such as PYPOWER \cite{PYPOWER} and pandapower
\cite{pandapower}. Users can and do report bugs by raising issues in
the GitHub repository \cite{PyPSA-github} or on the forum
\cite{PyPSA-forum}.

\begin{figure}[t]
  \includegraphics[trim=0.5cm 1cm 0.5cm 0.4cm,clip=true,width=\linewidth]{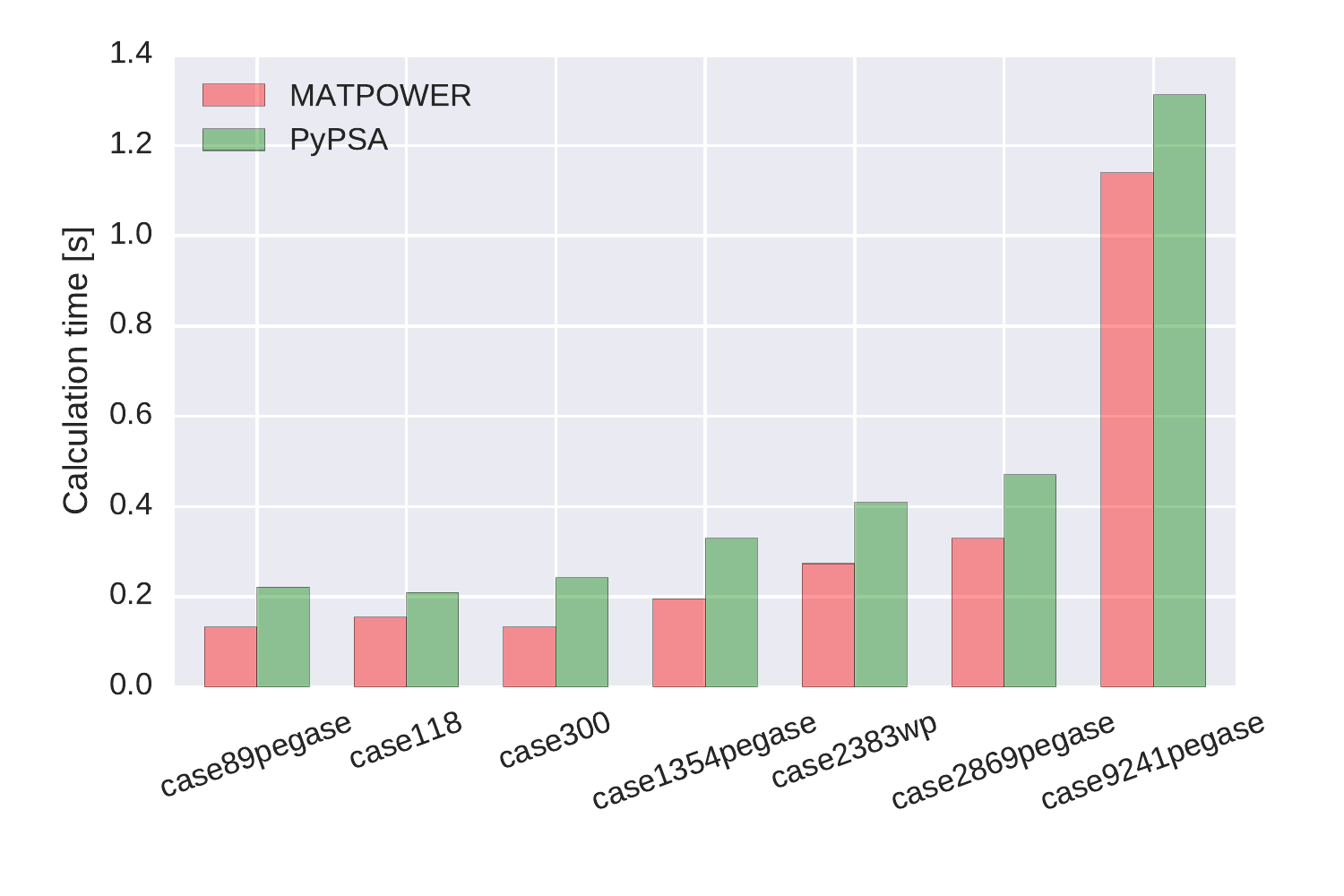}
  \caption{Calculation times for performing a full load flow on the MATPOWER \cite{MATPOWER} standard cases using MATPOWER versus PyPSA.}
\label{fig:pf_timing}
\end{figure}

\ifjors
\else
\section{Performance}\label{sec:performance}
\fi

In this section some examples of PyPSA's computational performance are given.

In Figure \ref{fig:pf_timing} computation times are given for a full
power flow on the MATPOWER \cite{MATPOWER} test cases (the IEEE
standard cases as well as snapshots from the French TSO RTE and
European networks \cite{RTE-PEGASE}) using MATPOWER and PyPSA. In both
cases the complete execution of the load flow function (`runpf' for
MATPOWER and `network.pf' for PyPSA) was timed on a computer with
Intel Core i5-2520M processors at 2.50GHz each with a tolerance of
$10^{-8}$ for the summed error in the apparent power $S$ from equation
\eqref{eq:pf}. The timings were averaged over 10 attempts for each
network. The computation times are similar, thanks to the fact that
both MATPOWER and PyPSA (via the SciPy library \cite{SciPy}) use the
same C library umfpack \cite{Davis:2004:AUV:992200.992206} for solving
sparse linear equation systems, but PyPSA is in all cases slightly
slower due to the overhead of preparing the admittance matrices in
pure Python code. If the admittance matrix remains the same for
several calculations, PyPSA has the option to avoid recalculating it,
which can save some of this time; further acceleration is possible by
using the just-in-time (jit) compiler numba \cite{numba}, as has been
done in the pandapower project \cite{pandapower} with success for
larger networks.

For the linear optimal power flow (LOPF) the computation performance
depends strongly on the choice of linear solver. To give an
indication of typical calculation times, if dispatch in the SciGRID
model of the German transmission network described in Section
\ref{sec:examples} (585 buses, 1423 generators including curtailable
wind and solar at each node, 38 pump storage units, 852 lines, 96
transformers) is optimised over 4 snapshots, it takes 5 seconds using
the COIN-OR Clp free solver on the computer described above. Extensive
timings for different formulations of the LOPF problem can be found in
\cite{2017arXiv170401881H}.


\ifjors
\else
\section{Comparison to other power system tools}\label{sec:comparison}
\fi

\ifjors
\begin{table}[t]
  \centering \footnotesize
  \setlength{\tabcolsep}{4.5pt}
\else
\begin{table*}[t]
  \centering
  \setlength{\tabcolsep}{6pt}
\fi
  \begin{tabular}{@{} clccc@{\hskip .7cm}c*{11}c @{}}
    & & & & & \multicolumn{3}{c}{Grid Analysis} & \phantom{} & \multicolumn{7}{c}{Economic Analysis} \\[2ex]
    \cmidrule{6-8}      \cmidrule{10-17}
    & Software & \rot{Version} & \rot{Citation} & \rot{Free Software} & \rot{Power Flow} & \rot{\shortstack[l]{Continuation\\Power Flow}}
    & \rot{\shortstack[l]{Dynamic\\Analysis}} & & \rot{Transport Model} & \rot{Linear OPF} & \rot{SCLOPF} & \rot{Nonlinear OPF}  & \rot{\shortstack[l]{Multi-Period\\Optimisation}}
    & \rot{Unit Commitment}  & \rot{\shortstack[l]{Investment\\Optimisation}}   & \rot{\shortstack[l]{Other Energy\\Sectors}} \\
    \cmidrule{2-17}
    & MATPOWER & 6.0 & \cite{MATPOWER} & \OK & \OK & \OK & & &\OK &  \OK &  &  \OK  \\
    & NEPLAN & 5.5.8 &\cite{NEPLAN}& & \OK &  & \OK & & \OK & \OK& \OK& \OK &&&&\OK\\
    & pandapower & 1.4.0 & \cite{pandapower} & \OK & \OK & & & &\OK &  \OK &  &  \OK \\
    & PowerFactory & 2017 & \cite{PowerFactory} & & \OK &  & \OK & &  & \OK & \OK & \OK  \\
    & PowerWorld & 19 & \cite{PowerWorld} & & \OK & & \OK & & \OK & \OK& \OK& \OK \\
    & PSAT & 2.1.10 & \cite{PSAT} & \OK & \OK & \OK & \OK & & & \OK &  & \OK & \OK & \OK & & \\
    & PSS/E & 33.10 & \cite{PSSE}& & \OK &  & \OK & &  & \OK & \OK & \OK  \\
    & PSS/SINCAL & 13.5 &\cite{SINCAL} & & \OK &  & \OK & &  & & & \OK &&&&\OK\\
    \rot{\rlap{~Power system tools}}
    & PYPOWER & 5.1.2 & \cite{PYPOWER} & \OK & \OK & & & &\OK &  \OK &  &  \OK \\
    \cmidrule{2-17}
    & PyPSA & 0.11.0 &  & \OK & \OK & & & & \OK & \OK & \OK & & \OK & \OK & \OK & \OK\\
    \cmidrule{2-17}
    & calliope & 0.5.2 &  \cite{Pfenninger20171}  & \OK & & & & & \OK & & & & \OK &  & \OK & \OK \\
    & minpower & 4.3.10 & \cite{minpower} & \OK  & & & & & \OK & \OK & & & \OK & \OK & &  \\
    & MOST & 6.0 & \cite{MOST}  & \OK & \OK & \OK & & & \OK &  \OK & \OK &  \OK & \OK & \OK \\
    & oemof & 0.1.4 & \cite{oemof} & \OK & & & & & \OK & & & & \OK & \OK & \OK & \OK \\
    & OSeMOSYS & 2017 & \cite{Howells20115850}  & \OK & & & & & \OK & & & & \OK &  & \OK & \OK \\
    & PLEXOS & 7.400 & \cite{PLEXOS} & & &  & & & \OK  & \OK & \OK & & \OK & \OK & \OK & \OK\\
    & PowerGAMA & 1.1 & \cite{doi:10.1063/1.4962415} & \OK  & &  & & & \OK  & \OK &  & & \OK \\
    & PRIMES & 2017 & \cite{Primes} & & &  & & & \OK  & \OK &  & & \OK & \OK & \OK & \OK\\
    & TIMES & 2017 & \cite{Times} & & &  & & & \OK  & \OK &  & & \OK & \OK & \OK & \OK\\
        \rot{\rlap{~Energy system tools}}
    & urbs & 0.7 & \cite{johannes_dorfner_2016_60484} & \OK & & & & & \OK & & & & \OK & \OK & \OK & \OK  \\
    \cmidrule[1pt]{2-17}
  \end{tabular}
  \caption{A comparison of selected features of selected software tools that are similar to PyPSA.}
  \label{tab:comparison}
\ifjors
\end{table}
\else
\end{table*}
\fi

Given the proliferation of software tools available for modelling
power systems, a guide is provided here that briefly compares PyPSA to
other power system tools, with a particular focus on free software in
the Python programming language. The advantages of Python are
discussed above in Section \ref{sec:software}.

Selected features for a selection of different software tools are
compared in Table \ref{tab:comparison}. Many of the tools have
specialised features that are not shown in the table, so this table
should only be treated as an indicative overview of their features in
relation to PyPSA's features.

Many power system tools concentrate on steady-state, dynamic
(i.e. short-term transient) and single-period OPF analysis of power
networks. They neglect the multi-period unit commitment, investment
optimisation and energy system coupling which PyPSA offers. In Python
we focus our comparison on two tools: PYPOWER and pandapower.

PYPOWER \cite{PYPOWER} is a port of an older version of MATPOWER
\cite{MATPOWER} from Matlab to Python. It does not make strong use of
Python's object-oriented interface and structures data using NumPy
arrays, which makes it difficult to track component attributes. It has
no functionality to deal with multi-period OPF, which makes it
unsuitable for unit commitment, storage optimisation or investment
optimisation. This reflects the functionality of older versions of
MATPOWER, but the latest version 6.0 of MATPOWER includes the MATPOWER
Optimal Scheduling Tool (MOST) \cite{MOST}, which does multi-period
OPF, but no investment optimisation. Unlike PyPSA, PYPOWER has the
ability to do full non-linear OPF for single snapshots.

pandapower \cite{pandapower} provides a pandas \cite{pandas} interface
to PYPOWER \cite{PYPOWER}, which makes it easier to use, and adds
useful functionality such as standard types (on which PyPSA's standard
types are based), short circuit calculations, state estimation,
and modelling of switches and three-winding transformers. The last four functions are
currently missing in PyPSA, along with non-linear OPF, but like PYPOWER,
pandapower does not have multi-period OPF functionality. pandapower is
under active development and the PyPSA team stays in contact with the
pandpower team to exchange tips and features, which is a clear benefit
for both developers and users of free software.

PyPSA differs from more general energy system models such as calliope
\cite{Pfenninger20171}, oemof \cite{oemof}, OSeMOSYS
\cite{Howells20115850} and urbs \cite{johannes_dorfner_2016_60484} by
offering more detailed modelling of power networks, in particular the
physics of power flow according to the impedances in the
network. PyPSA can model a more general energy network using link
components (see Section \ref{sec:coupling}), but cannot, for example,
yet do the multi-year dynamic investment that OSeMOSYS does. The
non-free PLEXOS software \cite{PLEXOS} comes the closest to matching
PyPSA's functionality, but PLEXOS is missing non-linear power flow.

These differences with other software tools are the reason that it was
decided to develop a new tool rather than to extend an existing
one. Existing tools for power flow such as PYPOWER did not have the
internal code and data structures for economic optimisation over
multiple time periods with many inter-temporal actors, whereas the
energy system tools were missing the tight integration with power flow
analysis that we believe is necessary for future research.

\ifjors
\else
\section{Demonstration of features on the SciGRID and GridKit datasets}\label{sec:examples}
\fi

On the PyPSA website \cite{PyPSA-website} a large number of examples of code
using PyPSA is linked for reference and to help users just starting out with the software. These range from basic small-scale networks demonstrating the features of PyPSA, to a one-node-per-country model of the European power system with high shares of renewables \cite{schlachtberger_david_2017_804338}, to full transmission network models available as open data from the SciGRID \cite{SciGRIDv0.2} and GridKit projects \cite{gridkit,Wiegmans} which we demonstrate here.

The SciGRID model of Germany provides geo-referenced data for substations and transmission lines (220~kV and above). In one code example, data from openly-available sources on power plant locations and capacities, load distribution and time series are added to the SciGRID data so that load flow calculations can be carried out. The results of one such simulation for Germany with nodal pricing is shown in Figure \ref{fig:scigrid}. In this snapshot there was a large amount of zero-marginal-cost wind feed-in suppressing the locational marginal prices ($\lambda_{n,t}$ from equation \eqref{eq:balance}) in the North of Germany. Transmission bottlenecks in the middle of Germany prevent the transportation of this cheap electricity to the South, where more expensive conventional generators set the price. The linearly-optimised dispatch was then fed into a full non-linear power flow calculation where each bus was set to maintain nominal voltage; the resulting reactive power feed-in is also shown in Figure \ref{fig:scigrid}.

The data quality for the transmission grid in OpenStreetMap outside
Germany is not of uniform quality, so for the European grid, an
extract of the ENTSO-E interactive map \cite{interactive} was made
\cite{gridkit} using GridKit \cite{Wiegmans}. The details of how load,
conventional power plants and renewable generation time series and
expansion potentials were added to the grid data are provided in a
forthcoming paper \cite{PyPSA-EU}. The result of generation and storage investment
optimisation for a clustering of the network from 5000 buses down to 256 buses,
allowing no grid expansion and assuming a CO2 reduction of 95\%
compared to 1990 levels, is shown in Figure \ref{fig:europe}. The lack
of grid expansion forces some balancing of renewable variability
locally with storage. Short-term battery storage (grey) combines with
solar power (yellow) in Southern Europe, while longer-term hydrogen
storage (purple) pairs with wind power (blue) in Northern Europe. This
system has an average cost of \euro~82/MWh. If the grid is optimally
expanded, much of the storage can be eliminated and costs are as low
as \euro~65/MWh \cite{Hoersch2017}.

\ifjors
\else
\section{Conclusions}\label{sec:conclusions}
\fi

In this paper a new toolbox has been presented for
simulating and optimising power systems. Python for Power System
Analysis (PyPSA) provides components to model variable renewable
generation, conventional power plants, storage units, coupling to
other energy sectors and multiply-connected AC and DC networks over
multiple periods for the optimisation of both operation and
investment. Tools are also provided for steady-state analysis with the
full load flow equations. PyPSA's performance for large datasets,
comparisons with other software packages and several example
applications are demonstrated.

As free software, the code of PyPSA can easily be inspected and
extended by users, thereby contributing to further research and also
transparency in power system modelling. Given that public acceptance
of new infrastructure is often low, it is hoped that transparent
modelling can contribute to public understanding of the various
options we face when designing a sustainable energy system.

\section*{Acknowledgments}

We thank Stefan Schramm for supporting the development of PyPSA. We thank
the community of PyPSA users for bug reports, improvement suggestions and their
general friendly support and encouragement for the further development of PyPSA.
This research was conducted as part of the CoNDyNet project, which is
supported by the German Federal Ministry of Education and Research
under grant no. 03SF0472C. The responsibility for the contents lies
solely with the authors.



%

\bibliographystyle{IEEEtran}
\bibliography{pypsa}

\end{document}